\documentclass[12pt,preprint]{aastex}
\slugcomment{Accepted to ApJ: 24 July 2014}

\usepackage{graphicx} 
\usepackage{natbib}
\usepackage{amsmath}
\usepackage{float}

\begin{document}

\title{Refined Rotational Period, Pole Solution \& Shape Model for (3200) Phaethon}

\author{
Megan Ansdell\altaffilmark{1,2},
Karen J.\ Meech\altaffilmark{1,2},
Olivier Hainaut\altaffilmark{3},
Marc W.\ Buie\altaffilmark{4},
Heather Kaluna\altaffilmark{1,2},
James Bauer\altaffilmark{5},
Luke Dundon\altaffilmark{6}
}

\altaffiltext{1}{NASA Astrobiology Institute}
\altaffiltext{2}{Institute for Astronomy, University of Hawaii, 2680 Woodlawn Drive, Honolulu, HI 96822, USA}
\altaffiltext{3}{European Southern Observatory, Karl Schwaszschildstrasse 2, 85748 Garching bei Muenchen, Germany}
\altaffiltext{4}{Southwest Research Institute, 1050 Walnut Street, Suite 300, Boulder, CO 80302, USA}
\altaffiltext{5}{Jet Propulsion Laboratry, 4800 Oask Grove Drive, MS 183-401, Pasadena, CA 91109}
\altaffiltext{6}{United States Navy, Washington, DC 20350, USA}

\email{mansdell@ifa.hawaii.edu}

\begin{abstract}

(3200) Phaethon exhibits both comet- and asteroid-like properties, suggesting it could be a rare transitional object such as a dormant comet or previously volatile-rich asteroid. This justifies detailed study of (3200) Phaethon's physical properties, as a better understanding of asteroid-comet transition objects can provide insight into minor body evolution. We therefore acquired time-series photometry of (3200) Phaethon over 15 nights from 1994 to 2013, primarily using the Tektronix 2048$\times$2048 pixel CCD on the University of Hawaii 2.2-m telescope. We utilized light curve inversion to: (1) refine (3200) Phaethon's rotational period to $P=3.6032\pm0.0008$ h; (2) estimate a rotational pole orientation of $\lambda=+85^{\circ}\pm13^{\circ}$ and $\beta=-20^{\circ}\pm10^{\circ}$; and (3) derive a shape model. We also used our extensive light curve dataset to estimate the slope parameter of (3200) Phaethon's phase curve as $G\sim0.06$, consistent with C-type asteroids. We discuss how this highly oblique pole orientation with a negative ecliptic latitude supports previous evidence for (3200) Phaethon's origin in the inner main asteroid belt as well as the potential for deeply buried volatiles fueling impulsive yet rare cometary outbursts. 

\end{abstract}

\keywords{minor planets: individual ((3200) Phaethon), techniques: photometric}

\newpage

\section{INTRODUCTION\label{sec-intro}}

(3200) Phaethon is a unique minor body in the Solar System, exhibiting both asteroid- and comet-like properties. For example, its unambiguous association with the Geminid meteor stream (\citealt{1983IAUC.3881....1W}; \citealt{1989A&A...225..533G}; \citealt{1993MNRAS.262..231W}) strongly supports a cometary origin, as meteor streams are typically formed from debris left along comet orbits as a result of volatile-driven activity. However, comet-like activity has never been observed in (3200) Phaethon (\citealt{2005ApJ...624.1093H}; \citealt{2008Icar..194..843W}), suggesting a more asteroid-like character (the recurrent dust tail at perihelion recently reported by \citealt{2013ApJ...771L..36J} is unlikely to result from volatile-driven activity). (3200) Phaethon also has features neither distinctly asteroidal nor cometary: its blue color fails to match the typical neutral/red colors of comet nuclei \citep{2004CometsII} while its spectral shape fails to closely match any meteoritic samples \citep{2007A&A...461..751L}.  (3200) Phaethon's orbit also brings it extremely close to the Sun, giving it an unusually small perihelion distance ($\sim$0.14 AU).

These atypical characteristics justify detailed study of (3200) Phaethon as it could be some form of rare transition object (e.g., an extinct/dormant comet or previously volatile-rich asteroid) that can provide insights into how minor bodies evolve. The combination of its size ($\sim$5.10 km), albedo ($\sim$0.11), and near-Earth orbit also allows (3200) Phaethon to periodically appear quite bright from Earth ($\sim$15 mag), permitting very precise ground-based observations.\footnote{Diameter and albedo from IRAS observations (IRAS-A-FPA-3-RDR-IMPS-V6.0)} Thus we have collected an extensive time-series photometry dataset for (3200) Phaethon and employed a light curve inversion technique (\citealt{2001Icar..153...37K}; \citealt{2001Icar..153...24K}) to refine the object's rotational period and pole orientation as well as derive a shape model. We focus on these key physical properties as they have important consequences for an object's origin and history (e.g., \citealt{2004Natur.428..400L}) as well as its thermophysical properties (e.g., \citealt{2009PASJ...61.1375O}). 

We begin in Section 2 by presenting the observations, data reduction, and time-series photometry that resulted in 16 light curves spanning almost two decades. In Section 3, we use this unprecedentedly large dataset for (3200) Phaethon to refine the object's rotational period as well as derive a highly oblique rotational pole with a negative ecliptic latitude. In Section 4 we discuss the implications of our findings, namely support for previous evidence of an origin in the inner main asteroid belt as well as the potential for deeply buried volatiles fueling impulsive yet rare cometary outbursts. We conclude in Section 5 with suggestions for future work to further investigate these possible scenarios.

\section{OBSERVATIONS, REDUCTION \& PHOTOMETRY \label{sec-reds}}
 
\subsection{Observations\label{sec-obs}}

We obtained time-series photometry over 15 nights from 1994 to 2013. These observations are summarized in Table~\ref{tab-obs}.  All but three nights used the Tektronix 2048$\times$2048 pixel CCD camera on the University of Hawaii 2.2-m telescope on Mauna Kea. Two nights used the PRISM 2048$\times$2048 pixel CCD camera on the Perkins 72-in telescope at the Lowell Observatory in Flagstaff, Arizona, while one night used the Optic 2048$\times$4096 CCD camera also on the University of Hawaii 2.2-m telescope. All observations used the standard Kron-Cousins {\it R} filter with the telescope guiding on (3200) Phaethon at non-sidereal rates.

\subsection{Reduction \& Photometry \label{sec-flat}}

Raw images were processed with standard IRAF routines for bias subtraction, flat-fielding, and cosmic ray removal \citep{1986SPIE..627..733T}.  We constructed reference flat-fields by median combining dithered images of either twilight or the object field (in both cases, flattening reduced gradients to $\textless$1\% across the CCD). We performed photometry using the IRAF $phot$ routine with circular apertures typically 5$''$ in radius, although aperture sizes changed depending on the night and/or exposure as they were chosen to consistently include 99.5\% of the object's light. Sky subtraction used either an annulus around the photometry aperture or median-combined samples of nearby patches of clear sky. (3200) Phaethon appeared point-like in all images, justifying our use of aperture photometry.

For photometric nights, we calibrated instrumental magnitudes using standard stars from \cite{1992AJ....104..340L}.  The standard stars ranged sufficiently in color and airmass to correct for color terms and extinction, thereby providing absolute flux calibrations. For non-photometric nights, the atmospheric extinction was typically only a few hundredths to tenths of a magnitude. Thus we calibrated these instrumental magnitudes using differential photometry with a large number (20-50) of field stars. Field stars were trailed in all images due to tracking on (3200) Phaethon at non-sidereal rates, so to perform photometry we used aperture sizes that included 99.5\% of a typical trailed field star's light for each night. When available, we used Sloan Digital Sky Survey (SDSS; \citealt{2000AJ....120.1579Y}) magnitudes to recover absolute flux calibrations; we computed transformations between Kron-Cousins and SDSS filters using equations provided on the SDSS website.\footnote{http://www.sdss.org} 

Unfortunately, SDSS did not cover the star fields of our Nov. 2004 and Dec. 2013 observing runs. For these nights, we used the weighted mean magnitude of each field star across stable periods (or over the entire night, if no period was sufficiently stable) as our reference magnitudes when performing differential photometry. These light curves should therefore be considered relative, rather than absolute. This approach was sufficient for the purposes of our analysis as the light curve inversion technique employed in this work can accommodate relative light curves.

\subsection{Lightcurves\label{sec-lcs}}

\begin{figure}
\centering
\includegraphics[width=17cm]{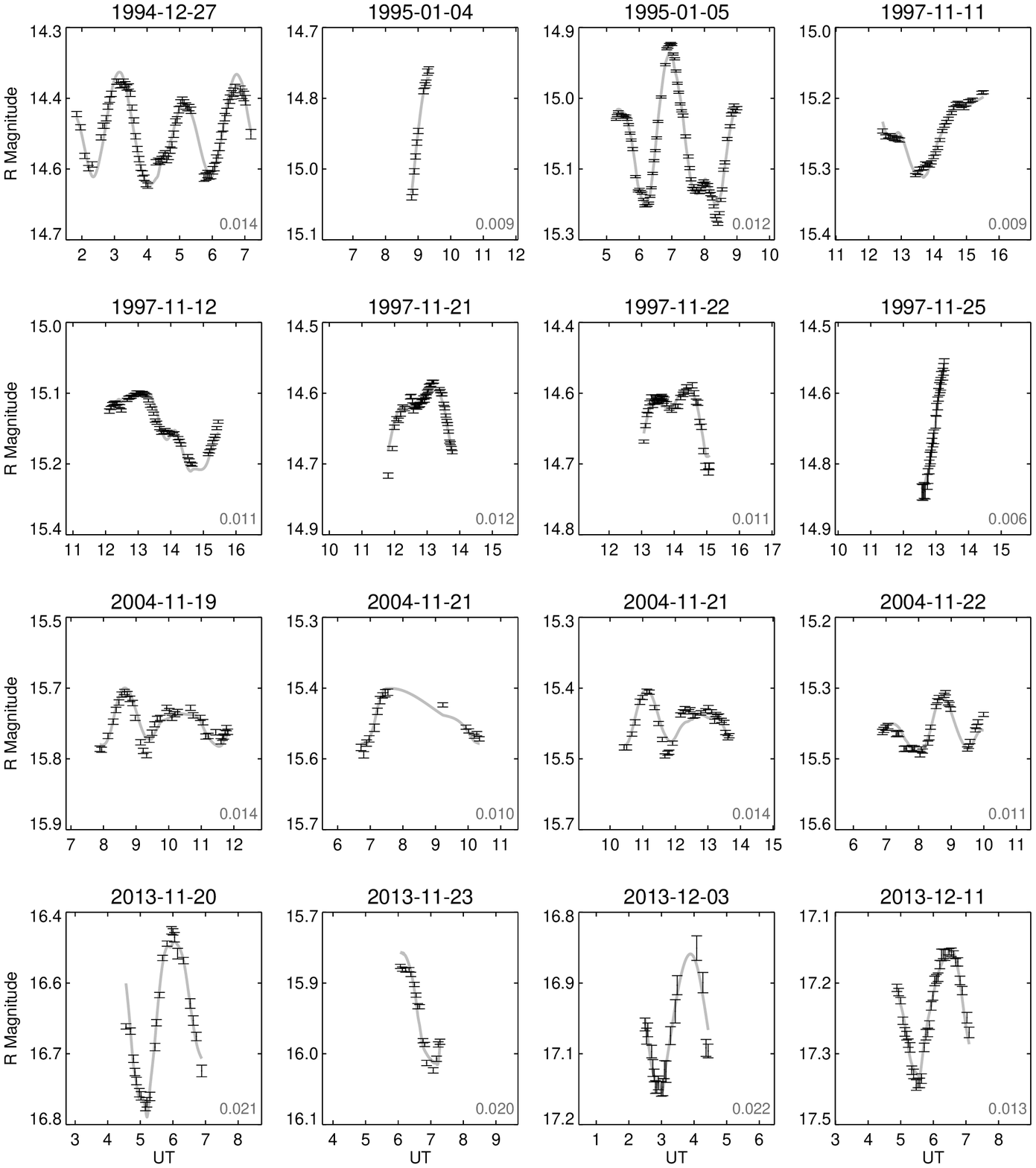}
\caption{\small Observed and model light curves of (3200) Phaethon. Observed light curves are shown in black with associated errors, while model light curves using our derived period and pole solutions are shown in gray. The RMS values between the observed and model light curves at each epoch are given in the lower right corners.}
\label{fig-lightcurve}
\end{figure}

Our observations resulted in 16 light curves, as shown in Figure~\ref{fig-lightcurve}. Note that one night (2004/11/21) contained two full light curves, thus 15 nights of observations resulted in 16 full or partial light curves. The supplementary online information contains a table of Julian Date (JD), Universal Time (UT), and {\it R} magnitude for each light curve point in our dataset. The light curves span many distinct viewing geometries (i.e., different combinations of ecliptic longitude and latitude) over roughly 20 years. On average, they cover $\sim$75\% of a full rotation period with $\sim$33 photometry points. Their asymmetric double peaks and significant changes in shape and amplitude over time are probably due to a combination of changing viewing geometries as well as (3200) Phaethon's high orbital inclination relative to the ecliptic  ($\sim$$22^\circ$) and its potentially non-spherical shape (see Figure~\ref{fig-shape}).

\subsection{Phase Curve\label{sec-lcs}}

\begin{figure}
\centering
\includegraphics[width=7.5cm]{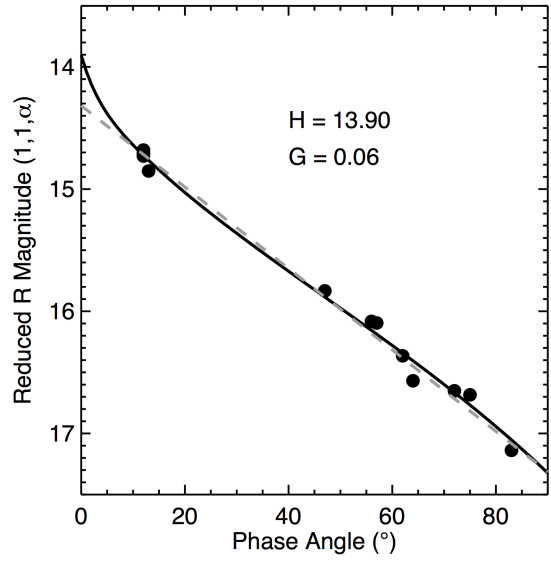}
\caption{\small Phase curve of (3200) Phaethon. The gray dashed line is the linear least-squares fit to the black data points (errors on the data are smaller than the symbols). The black curve is the best-fit phase curve model using the HG formalism, where $H=13.90$ and $G=0.06$.}
\label{fig-phase}
\end{figure}

Our dataset spans a wide range of phase angles ($\alpha$), allowing us to construct a phase curve for (3200) Phaethon. For each night with absolute calibrations, we calculated the reduced {\it R} magnitude using the standard equation:

\begin{equation}
H(\alpha) = m(\alpha)_{R} - 5log({\rm R}\Delta),
\end{equation}

\noindent where $m(\alpha)_{R}$ is the weighted mean of the observed {\it R} magnitudes at a given $\alpha$, and R and $\Delta$ are the associated heliocentric and geocentric distances in AU, respectively. As shown in Figure~\ref{fig-phase}, our dataset only covers the linear portion of (3200) Phaethon's phase curve. Thus we could strongly constrain \textit{G} (the slope parameter), but not \textit{H} (the absolute magnitude), as observations at small phase angles are important for constraining the upturn of the model phase function when using the HG formalism \citep{1989aste.conf..524B}. We minimized the residuals between the data and the model phase function, resulting best-fit model parameters of $H=13.90$ and $G=0.06$. This low value of \textit{G} is consistent with C-type asteroids, which have typical \textit{G} values of 0.05$\pm$0.02 (compared to S-type asteroids with typical \textit{G} values of 0.23$\pm$0.02; \citealt{1990A&AS...86..119L}). This agrees with (3200) Phaethon's classification as an F-type \citep{1985IAUC.4034....2T} or B-type \citep{1985MNRAS.214P..29G} asteroid, both subtypes of C-type asteroids.

\section{PERIOD, POLE ORIENTATION  \& SHAPE MODEL \label{sec-analysis}}

\subsection{Light Curve Inversion \label{sec-lcinversion}}

Light curve inversion is used to derive rotational states and shape models from disk-integrated, time-series photometry. Light curves represent the instantaneous scattered sunlight received at Earth from the projected surface area of a rotating object. The projected surface changes with viewing geometry, affecting the observed light curve amplitude and shape. With data at a sufficient number of different viewing geometries (i.e., different combinations of ecliptic longitude and latitude), it is possible to reconstruct the unprojected shape of the object and solve for its rotational state.

We use the light curve inversion software \textit{convexinv} to refine the rotational period and pole orientation of (3200) Phaethon as well as determine its shape. \textit{Convexinv} uses the light curve inversion scheme described in \cite{2001Icar..153...37K} and \cite{2001Icar..153...24K} to compute a shape-spin-scattering model that gives the best fit to a set of input light curves. The software is available online at the Database of Asteroid Models from Inversion Techniques (DAMIT; \citealt{2010A&A...513A..46D}).\footnote{http://astro.troja.mff.cuni.cz/projects/asteroids3D/web.php}

\subsection{Rotational Period \label{sec-period}}

Previous estimates of (3200) Phaethon's rotational period used only 2--3 light curves and showed significant spread, from 3.57$\pm$0.02h \citep{1998Icar..136..124P} to 3.604$\pm$0.001 h \citep{1996ACM}. We therefore refined (3200) Phaethon's rotational period by inputing our dataset into the \textit{period$\_$scan} program (part of the \textit{convexinv} package). \textit{Period$\_$scan} searches a user-specified period interval to find the best-fit model to the input light curves, as defined by a minimum relative $\chi$$^2$ value ($\chi$$_{\rm Rel}^2$; see \textit{convexinv} documentation for details). 

\begin{figure}
\centering
\includegraphics[width=8.6cm]{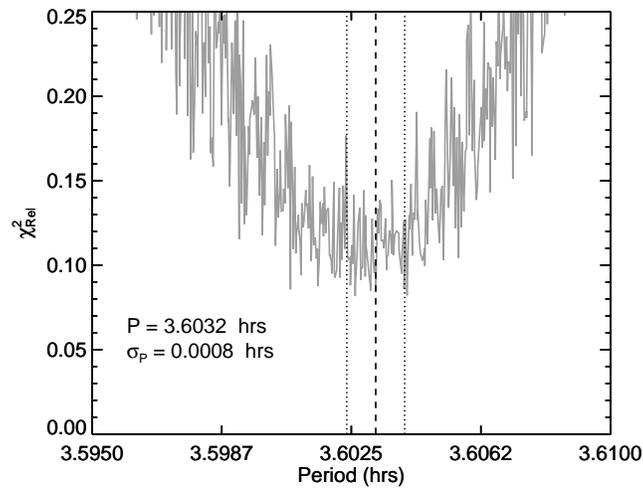}
\caption{\small $\chi$$_{\rm Rel}^2$ minimization plot for the rotational period of (3200) Phaethon output from  {\it period\_scan}. Dashed line shows our final period estimate of $P=3.6032$ h, while dotted lines show our estimated errors of $\sigma_{P}=0.0008$ h. Because {\it period\_scan} does not consider observational errors, we used a Monte Carlo method to determine these values (see \S\ref{sec-period}).}
\smallskip
\label{fig-periodscan}
\end{figure}

We first scanned 1.0--10.0 h to confirm the global $\chi$$_{\rm Rel}^2$ minimum near $\sim$3.6 h, then scanned 3.595--3.610 h at intervals of 3$\times$10$^{-5}$ h to refine this period; Figure~\ref{fig-periodscan} shows the resulting $\chi$$_{\rm Rel}^2$ minimization plot. Because \textit{period$\_$scan} does not take into account observational errors, we determined the best-fit period and associated error using a Monte Carlo approach. We added random Gaussian-distributed noise scaled to typical photometry errors ($\sim$$0.01$ mags) to each light curve point, then used \textit{period$\_$scan} to find the period associated with the minimum $\chi$$_{\rm Rel}^2$ value. We repeated this 100 times, taking the mean and standard deviation of the results as our final period estimate, $P=3.6032\pm0.0008$ h.

\subsection{Pole Orientation\label{sec-pole}}

\begin{figure*}
\centering
\includegraphics[width=10cm]{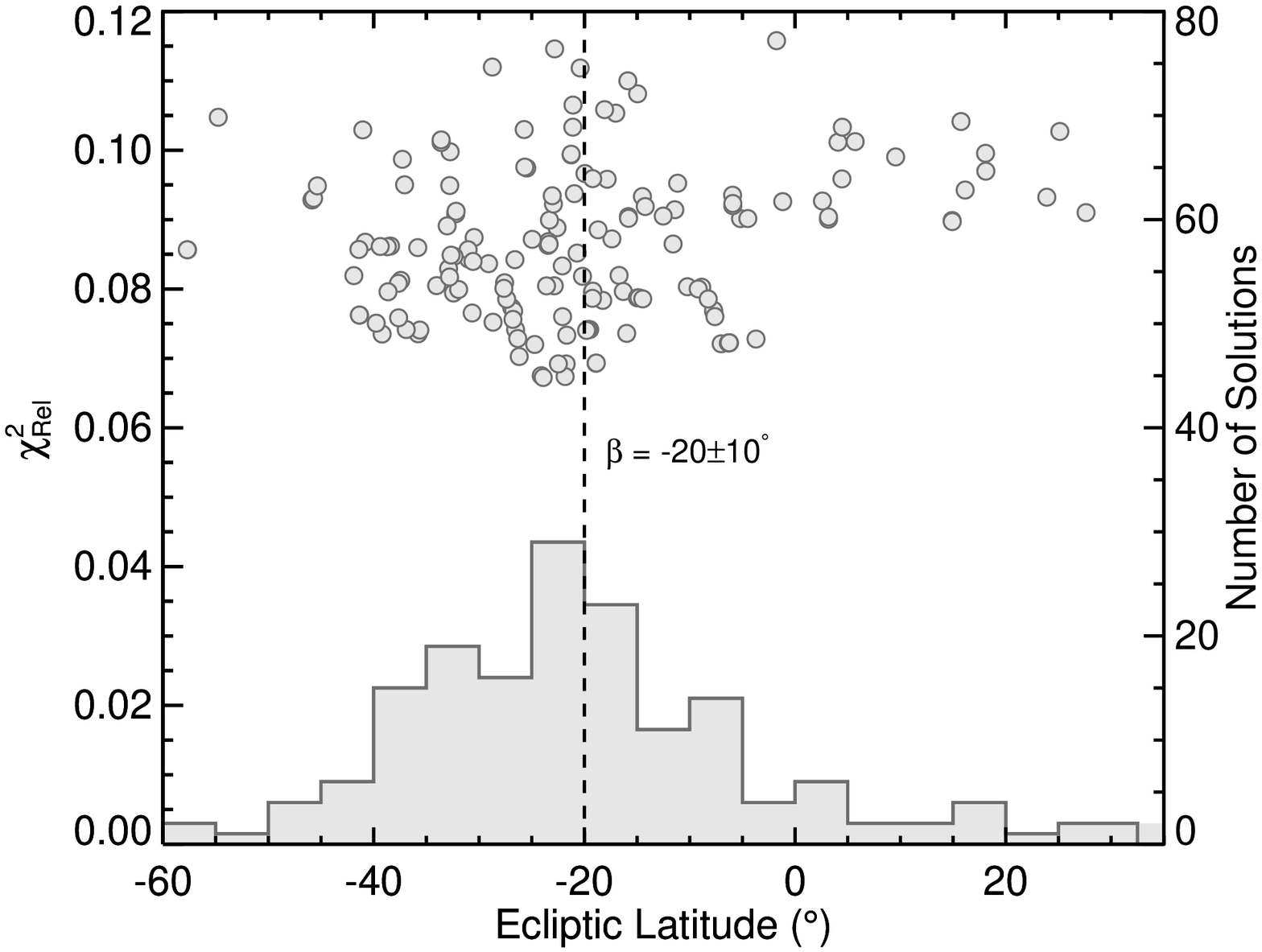}
\caption{\small Best-fit ecliptic latitude solutions output from \textit{convexinv} across our range of possible periods. The histogram shows clear clustering of best-fit solutions at $\beta$$\sim$--20$^{\circ}$, while the $\chi$$_{\rm Rel}^2$ points show a trend toward minimum values also at $\beta$$\sim$--20$^{\circ}$. The final pole solution, $\beta=-20\pm10^{\circ}$ (dashed line), was found using the mean and standard deviation of all solutions within 10\% of the lowest $\chi$$_{\rm Rel}^2$ value.}
\smallskip
\label{fig-lat}
\end{figure*}

\begin{figure*}
\centering
\includegraphics[width=10cm]{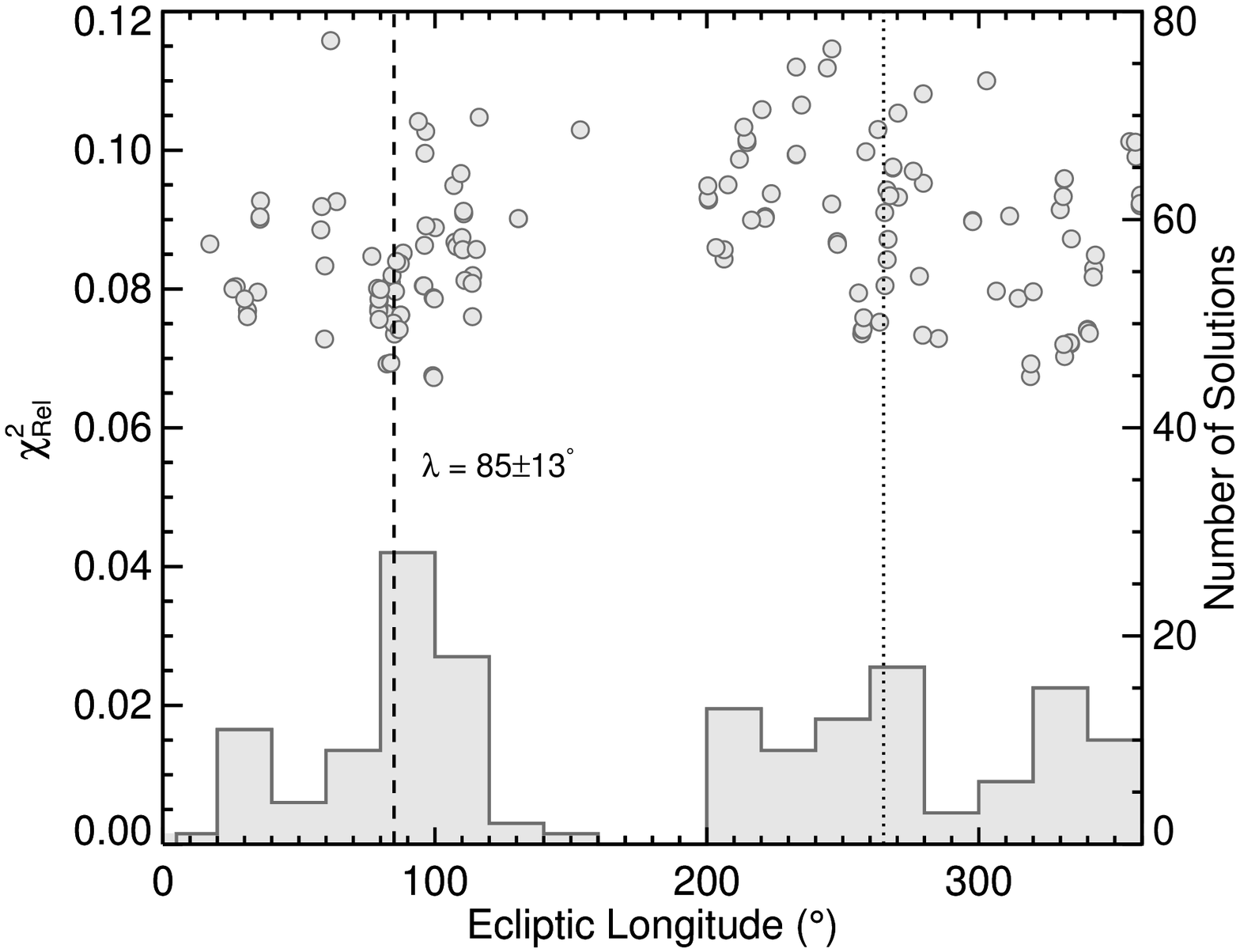}
\caption{\small Best-fit ecliptic longitude solutions output from \textit{convexinv} across our range of possible periods. The histogram shows clear clustering of best-fit solutions at $\lambda$$\sim$90$^{\circ}$, while the $\chi$$_{\rm Rel}^2$ points show a trend toward minimum values also at $\lambda$$\sim$90$^{\circ}$. The final pole solution, $\lambda=85\pm13^{\circ}$ (dashed line), was found using the mean and standard deviation of all solutions within 10\% of the lowest $\chi$$_{\rm Rel}^2$ value (after rejecting all solutions more than $\pm90^{\circ}$ from $\lambda$$\sim$$90^{\circ}$ to avoid contamination from possible mirror solutions). The dotted line shows a possible 180$^{\circ}$ mirror solution.}
\smallskip
\label{fig-long}
\end{figure*}

\begin{figure*}
\includegraphics[width=18cm]{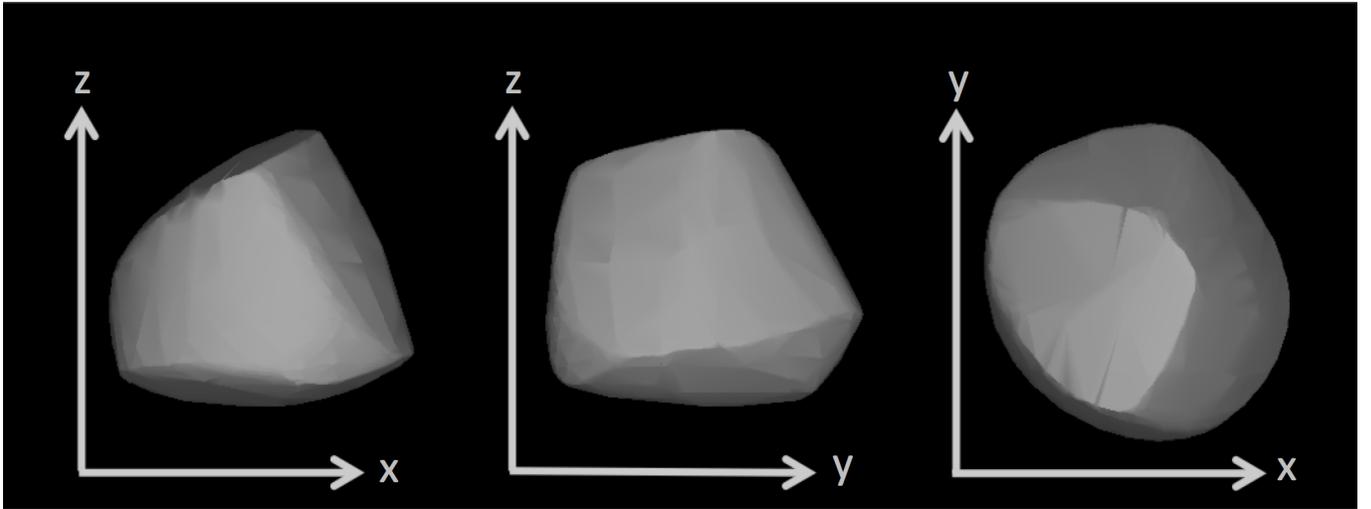}
\caption{\small Shape model of (3200) Phaethon derived using our best-fit period and pole solutions. The model is shown in three orthogonal views: the left and center panels show equatorial views that are 90$^{\circ}$ apart (pole oriented upwards) while the right panel shows a pole-on view (pole oriented out of the page). The axis ratios are $x/y\approx1.04$ and $x/z\approx1.14$.}
\smallskip
\label{fig-shape}
\end{figure*}

To derive pole solutions {\it convexinv} requires a set of light curves, an estimated rotational period, and an initial guess of pole orientation. Given these inputs the program performs a user-specified number of iterations until converging on a best-fit pole solution defined by a minimum $\chi$$_{\rm Rel}^2$ value. To find our best-fit pole solution given our uncertainties on (3200) Phaethon's rotational period, we performed 160 ``runs''  where each run used a unique period (covering the range found in \S\ref{sec-period} with a resolution of 1$\times$10$^{-5}$ h) and tested a grid of 156 initial pole guesses (equally spaced in ecliptic coordinate space). We took the best-fit solution from each run as that associated with the minimum $\chi$$_{\rm Rel}^2$ value. The best-fit results from each of these 160 runs are shown in Figure~\ref{fig-lat} (for ecliptic latitude) and Figure~\ref{fig-long} (for ecliptic longitude); there is clear clustering of best-fit results at $\beta$$\sim$$-20^{\circ}$ and $\lambda$$\sim$$+85^{\circ}$. We determined our final pole solution by taking the mean and standard deviation of the results within 10\% of the lowest $\chi$$_{\rm Rel}^2$ value across all runs (in order to filter out poor fits). We also omitted solutions more than $\pm90^{\circ}$ from $\lambda$$\sim$$+85^{\circ}$ in order to avoid contamination from possible mirror solutions (due to the 180$^{\circ}$ degeneracy in $\lambda$ that is common when using light curve inversion to derive rotational pole orientations). This gave a final pole solution of $\lambda=+85^{\circ}\pm13^{\circ}$ and $\beta=-20^{\circ}\pm10^{\circ}$.

Our results confirm one of the preliminary pole solutions found by \cite{2002Icar..158..294K}, namely $\lambda$$_1$ = 97$^{\circ}$$\pm$15$^{\circ}$ and $\beta$$_1$ = --11$^{\circ}$$\pm$15$^{\circ}$. Thus (3200) Phaethon appears to have a highly oblique rotational pole. Obliquity is the angle between an object's rotational pole and the normal to its orbital plane, where high obliquity refers to a pole oriented very close to the orbital plane. (3200) Phaethon's high orbital inclination of $i\approx22^{\circ}$ relative to the ecliptic, combined with its longitude of ascending node at $\Omega\approx256^{\circ}$, means that the rotational pole derived above could be only $\sim$2$^{\circ}$ above its orbital plane, corresponding to a notably high obliquity of $\sim$88$^{\circ}$.

\subsection{Shape Model} \label{sec-shape}

\textit{Convexinv} uses a set of input light curves and a user-defined rotation period and pole orientation to derive a shape model in the form of a convex polyhedron described by a set of triangular facets and vertices \citep{2001Icar..153...24K}. Prior to this work, a shape model for (3200) Phaethon had not been attempted because of insufficient data and thus uncertain period and pole estimates. Figure~\ref{fig-shape} presents our shape model for (3200) Phaethon using the rotational period and pole orientation derived above; the axis ratios are $x/y\approx1.04$ and $x/z\approx1.14$. We performed a sensitivity analysis by testing period and pole solutions randomly perturbed by our estimated errors. We found that in some cases 3200 Phaethon was predicted to be a long-axis rotator (i.e., $x/z<1$). This is an interesting result, as long-axis rotation indicates a perturbed state (e.g., \citealt{2004come.book..281S}). However when visually comparing the model results to the data, our best-fit solutions gave a substantially better fit to the data (see Figure~\ref{fig-lightcurve}) than the perturbed solutions.

\section{DISCUSSION\label{sec-discussion}}

\subsection{Preferential Heating at Perihelion \label{sec-heating}}

\begin{figure}
\centering
\includegraphics[width=8.cm]{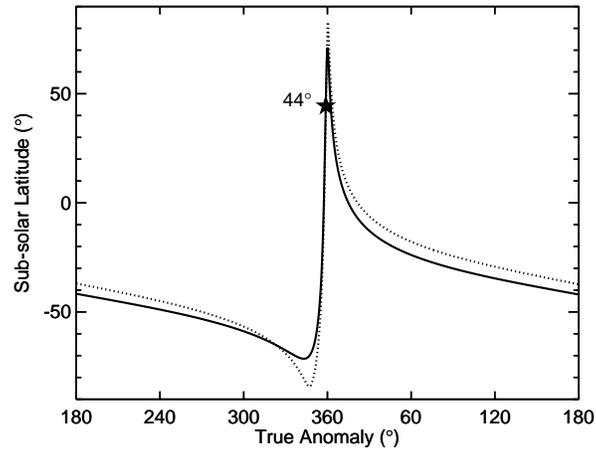}
\caption{\small Variations in sub-solar latitude along (3200) Phaethon's orbit. The star indicates perihelion, when the sub-solar point on (3200) Phaethon is at $\sim$44$^{\circ}$ latitude. The solid line shows the sub-solar latitude when using the pole orientation derived in this work. The dotted line shows the sub-solar latitude when using the pole orientation derived by \cite{2002Icar..158..294K} and used in the thermal analysis of \cite{2009PASJ...61.1375O}. Both indicate that (3200) Phaethon experiences preferential heating of its northern hemisphere at perihelion.}
\smallskip
\label{fig-sslat}
\end{figure}

\cite{2009PASJ...61.1375O} performed a detailed thermal analysis of (3200) Phaethon using the preliminary rotational pole initially found by \cite{2002Icar..158..294K} and confirmed in this work. They found that (3200) Phaethon's highly oblique rotational pole, combined with its highly inclined and eccentric orbit, causes preferential heating of its northern hemisphere at perihelion. This is illustrated in Figure~\ref{fig-sslat} (to be compared to Figure 1b in \citealt{2009PASJ...61.1375O}), which shows (3200) Phaethon's sub-solar latitude as a function of true anomaly. When (3200) Phaethon is at perihelion (i.e., only 0.14 AU from the Sun) its northern hemisphere appears to be exposed to intense heating. 

\cite{2009PASJ...61.1375O} also calculated (3200) Phaethon's sub-solar equilibrium surface temperature as a function of its orbit (see their Figure 2). They found that at perihelion solar-radiation heating causes the surface temperature of (3200) Phaethon to exceed 800 K, sufficient to decompose and dehydrate minerals such as serpentine phyllosilicates. \cite{2009PASJ...61.1375O} hypothesized that this could result in latitude-dependent color variations on the surface of (3200) Phaethon. Preferential heating would thermally metamorphose and dehydrate phyllosilicates in the more exposed areas, altering the mineralogy to create a visibly bluer surface in the northern hemisphere. They then used previously published spectra to test whether (3200) Phaethon's spectral gradient became more negative (i.e., bluer) as the sub-Earth point approached the object's north pole and thus brought more of the thermally metamorphosed areas into view for an observer on Earth. Although their results hinted at a bluer surface at northern latitudes, they were ultimately inconclusive.

We therefore searched for color variation as a function of sub-Earth latitude using nights from our dataset that contained multi-filter ({\it BVR}) photometry supplemented with literature color values (see Table~\ref{tab-color}).  Although none of these observations cover positive sub-Earth latitudes, the northern hemisphere should be increasingly visible for sub-Earth points $>-45^{\circ}$. We calculated our $B-V$ and $V-R$ colors by interpolating our {\it R} magnitudes to the UT times of the other filter measurements, using propagation of errors for subtraction to estimate uncertainties. Due to the fine time sampling of our {\it R}-band data (typically $\leq$ 2 min), uncertainties on the interpolated magnitudes are probably not larger than for the individual measurements. As shown in Table~\ref{tab-color}, we found that $B-V$ color for (3200) Phaethon does decrease (become bluer) as the sub-Earth point approaches the northern hemisphere. Although $V-R$ color does not show a clear trend, these filters sample redder wavelengths. These results therefore support \cite{2009PASJ...61.1375O}'s prediction of preferential thermal processing in the northern hemisphere of (3200) Phaethon at perihelion.

It is important to note that the color variation predicted by \cite{2009PASJ...61.1375O} requires that the preferential heating of the present planetary epoch be the primary metamorphic heat source of (3200) Phaethon. In other words, (3200) Phaethon could not have been heated to more than a few hundred degrees prior to being injected into its current near-Sun orbit roughly $\sim$10$^3$ years ago (corresponding to the age of the Geminids; \citealt{1989A&A...225..533G}, \citealt{2007MNRAS.375.1371R}). This is plausible as (3200) Phaethon is classified as an F- or B-type asteroid; these asteroids have been associated with CI/CM carbonaceous chondrites \citep{1996M&PS...31..321H}, which are believed to have undergone only moderate heating that can lead to aqueous alteration but not thermal metamorphism.

\subsection{Volatile Survival} \label{sec-volatiles}

The evidence for preferential heating of (3200) Phaethon's northern hemisphere at perihelion, discussed above, raises the possibility of deeply buried volatiles surviving despite an extremely close approach to the Sun. Although a previous calculation of (3200) Phaethon's core temperature at $\sim$250 K \citep{2005ApJ...624.1093H} is too high for water ice to survive, this estimate assumed thermal equilibrium. Extreme pole orientations, such as the one found in this work, may allow cooler core temperatures because thermal equilibrium would no longer apply. \cite{2013DPS....4541332B} performed a more detailed three-dimensional ``physico-chemical" modeling of (3200) Phaethon using a highly oblique pole similar to the one found in this work in order to assess whether water ice could still exist in the core of (3200) Phaethon. They found that (3200) Phaethon is likely to contain relatively pristine volatiles in its interior despite repeated close approaches to the Sun, leaving open the possibility of impulsive outbursts as deeply buried volatiles break through the volatile-depleted surface layers. Thus previous failed attempts to detect comet-like activity on (3200) Phaethon (\citealt{2005ApJ...624.1093H}; \citealt{2008Icar..194..843W}) may have been unsuccessful simply because the observations did not coincide with an outburst. 

Although very small amounts of activity at perihelion have been observed in (3200) Phaethon using the space-based STEREO solar observatory, this activity was interpreted to arise from thermal fracture and desiccation cracking due to intense heating at perihelion, rather than comet-like activity from deeply buried volatiles \citep{2010AJ....140.1519J}. Moreover, the mass-loss rate from the short-lived dust tails at perihelion are insufficient to account for ongoing replenishment of the Geminid stream, and the dust particles may also be gravitationally unbound to the Solar System, preventing them from contributing to the Geminids \citep{2013ApJ...771L..36J}.

\subsection{Main Belt Origin\label{sec-discusspole}}

(3200) Phaethon has been linked to the main asteroid belt in previous dynamical and compositional studies. \cite{2002Icar..156..399B} used dynamical modeling of (3200) Phaethon's orbit to show that it has a zero probability of originating from comet reservoirs such as the Jupiter Family Comet region, but a 50\% and 80\% probability of originating from the central and inner main asteroid belt, respectively. \cite{2010A&A...513A..26D} then made the compositional link by showing significant similarities between the reflectance spectra of (3200) Phaethon and another B-type asteroid in the central main belt, 2 Pallas. 

The negative ecliptic latitude of (3200) Phaethon's pole derived in this work supports its origin in the inner main asteroid belt. Objects in near-Earth space have a distinct excess of retrograde spins \citep{2007Icar..192..223K}, which has been postulated to result from the dynamical mechanism that transfers objects from the inner main asteroid belt into near-Earth space---namely the highly efficient $\nu$$_6$ resonance. Because the $\nu$$_6$ resonance is located at the inner edge of the main belt, it can only be reached by asteroids with orbits evolving inward toward the Sun. The Yarkovsky effect is the well-known mechanism that alters the orbital semi-major axes of asteroids (see \citealt{2002aste.conf..395B} for an overview of the influence of the Yarkovsky effect on the dynamical evolution of asteroids), however the Yarkovsky effect only evolves orbits inward for asteroids with retrograde rotations (i.e., $\beta<0$). Therefore the observed excess of retrograde spins among near-Earth objects has been explained by ``dynamical filtering'' when retrograde main belt asteroids evolving inward due to the Yarkovsky effect are preferentially ejected into near-Earth space via the $\nu$$_6$ resonance \citep{2004Natur.428..400L}.  
 
It is important to note that the Yarkovsky effect is less efficient for objects with highly oblique poles, such as (3200) Phaethon. Therefore we must consider alternative mechanisms for altering (3200) Phaethon's rotational pole to such an extreme orientation while in near-Earth orbit. Collisions can alter rotational pole orientations, although they are highly improbable for an object like (3200) Phaethon due to its small size and the limited population of potential impactors in the inner Solar System. The Yarkovsky-O'Keefe-Radzievskii-Paddack (YORP) effect can also affect spin states, however timescales for changing obliquity by $\sim$90$^{\circ}$ are typically million years \citep{2000Icar..148....2R}; given the young age of the Geminids ($\sim$10$^3$ years; \citealt{1989A&A...225..533G}; \citealt{2007MNRAS.375.1371R}), it is unlikely that there has been sufficient time for the YORP effect to significantly alter (3200) Phaethon's rotational state. However, if (3200) Phaethon is indeed still active, variable outbursts may potentially explain the object's extreme pole orientation.

\section{SUMMARY \& FUTURE WORK\label{sec-summarywork}}

We have used an extensive time-series photometry dataset, consisting of 16 light curves spanning over 20 years, to refine (3200) Phaethon's rotational period and pole orientation as well as derive its shape model. We find a period of $P=3.6032\pm0.0008$ h with a pole orientation of $\lambda=+85^{\circ}\pm13^{\circ}$ and $\beta=-20^{\circ}\pm10^{\circ}$. Key areas of future work include confirming surface color variation due to preferential heating at perihelion (e.g., by measuring $B-V$ and $V-R$ colors at positive sub-Earth latitudes) and continuing the search for low-level cometary outbursts (e.g., by serendipitous observation).  

Another important area of future work will be deriving pole solutions for the two smaller minor bodies associated with (3200) Phaethon---2005 UD  \citep{2006AJ....132.1624J} and 1999 YC \citep{2008M&PSA..43.5055O}---in order to assess their possible formation mechanisms. If all three bodies have randomized pole orientations, this may point to their formation via explosive activity in (3200) Phaethon soon after it was transferred to its current near-Sun orbit from the main belt. Suddenly exposing a volatile-rich (3200) Phaethon to intense heating at perihelion could have resulted in a burst of activity that formed the Geminids, as well as 2005 UD and 1999 YC,  leaving (3200) Phaethon dormant/extinct. This is an enticing interpretation given that the Geminids is a dynamically young meteor stream, which suggests ongoing activity, yet (3200) Phaethon has exhibited no known comet-like outbursts sufficient to replenish the stream. Simultaneous formation of the Geminids, 2005 UD, and 1999 YC with extinction/dormancy of (3200) Phaethon would account for both the youth of the Geminids as well as the lack of activity seen in (3200) Phaethon. However because such an event could significantly alter rotational states, the negative ecliptic latitude of (3200) Phaethon's pole would no longer be evidence for its origin in the inner main belt (although this would not preclude a possible origin in the main belt).

\begin{acknowledgements}
We thank the referee for their very detailed review of this work, which greatly improved the paper. We are extremely grateful to the telescope staff of UH88 and Lowell Observatory, without whom this work would not have been possible. A special thanks to Michael Belton for his insightful feedback on short notice. We also thank Josef \v{D}urech and his collaborators, who kindly provided the MATLAB code used to generate the shape model visualization. This work was supported by the National Aeronautics and Space Administration through the NASA Astrobiology Institute under Cooperative Agreement No. NNA04CC08A issued through the Office of Space Science, by NASA Grant Nos. NAGW 5015, NAG5-4495, NNX07A044G, NNX13A151G, and NNX07AF79G. Image processing was done in part using the IRAF software. IRAF is distributed by the National Optical Astronomy Observatories, which is operated by the Association of Universities for Research in Astronomy, Inc. (AURA) under cooperative agreement with the National Science Foundation.
\end{acknowledgements}



\clearpage

\clearpage

\clearpage

\clearpage

\clearpage

\clearpage

\clearpage


\clearpage
\begin{deluxetable}{p{1.0cm}lllcccccccccccc}
\tabletypesize{\scriptsize}
\rotate
\tablewidth{0pt}
\tablecaption{(3200) Phaethon Dataset \label{tab-obs}}
\tablecolumns{15}
\tablehead{
   \colhead{Obs. Date (UT)}
  & \colhead{Instrument}
 & \colhead{Observer(s)}
 & \colhead{Weather}
 & \colhead{Seeing (\textsuperscript{${\prime}$${\prime}$})}
 & \colhead{$\#$ Obs.\textsuperscript{a}}
 & \colhead{Expt.\textsuperscript{b}}
 & \colhead{$\%$ Rot.\textsuperscript{c}}
 & \colhead{RN\textsuperscript{d}}
 & \colhead{Gn\textsuperscript{e}}
 & \colhead{$\Delta$}
 & \colhead{R}
 & \colhead{$\lambda$}
 & \colhead{$\beta$}
 & \colhead{$\alpha$}
}
\startdata
1994/12/27&PRISM&Buie&Photometric& 1.6 & 76 & 13680 &  149 & 7.2 & 2.7 & 0.44 & 1.22 &  28 & 10 & 47 \\
1995/01/04&Tektronix&Meech  \& Hainaut&Photometric& 0.6 & 11 & 1100 & 15 & 10.0 & 1.8 & 0.46 &  1.11 &  16 & 6 & 62 \\
1995/01/05&Tektronix&Meech  \& Hainaut&Photometric& 0.7 & 79 & 7860 & 103 & 10.0 & 1.8 & 0.46 & 1.09 &  15 & 6 & 64 \\
1997/11/11&Tektronix&Meech  \& Bauer&Photometric& 1.3 & 39 & 5850 &  86 & 6.0 & 1.8 & 0.57 &  1.18 & 133 & 7 & 56 \\
1997/11/12&Tektronix&Meech  \& Bauer&Photometric& 1.2 & 52 & 6240 &  93 & 6.0 & 1.8 & 0.55 &  1.17 & 135 & 7 & 57 \\
1997/11/21&Tektronix&Meech  \& Bauer&Photometric& 1.0 & 48 & 2880&  55 & 6.0 & 1.8 & 0.38 &  1.03 & 153 & 5 & 72 \\
1997/11/22&Tektronix&Meech  \& Bauer&Photometric& 0.8 & 41 & 1260 &  81 & 6.0 & 1.8 & 0.38 &  1.01 & 156 & 4 & 75  \\
1997/11/25&Tektronix&Meech  \& Bauer&Photometric& 0.5 & 24 & 720 &  19 & 6.0 & 1.8 & 0.34 &  0.96 & 166 & 3 & 83 \\
2004/11/19&Tektronix&Dundon&Cirrus& 0.9 & 38 & 1860 &  109 & 6.0 & 1.8 & 0.84 & 1.78 &  76 & 17 & 13 \\
2004/11/21a&Tektronix&Dundon&Cirrus& 1.0 & 16 & 640 &  101 & 6.0 & 1.8 & 0.81 & 1.76 &  75 & 17 & 12 \\
2004/11/21b&Tektronix&Dundon&Cirrus& 1.0 & 35 & 1400 &  90 & 6.0 & 1.8 & 0.81 &  1.76 &  75 & 17 & 12 \\
2004/11/22&Tektronix&Dundon& Cirrus & 1.1 & 35 & 1400 &  86 & 6.0 & 1.8 & 0.80 &  1.75 &  74 & 17 & 12 \\
2013/11/20&Tektronix&Dundon& Cirrus & 0.9 & 24 & 2880 &  65 & 14.0 & 1.8 & 0.79 &  1.07 &  308 & 28 & 62 \\
2013/11/23&Optic&Ansdell&Cirrus& 0.9 & 16 & 2400 &  34 & 4.0 & 1.4 & 0.84 &  1.12 &  314 & 28 & 58 \\
2013/12/03&PRISM& Meech \& Ansdell& Cirrus & 2.4 & 20 & 3030 &  54 & 7.2 & 2.7 & 1.02 & 1.26 & 328  & 25 & 50 \\
2013/12/11&Tektronix&Ansdell& Photometric & 1.3 & 26 & 3900 &  62 & 14.0 & 1.8 & 1.17 &  1.37 &  336 & 23 & 45 \\
\enddata
\tablenotetext{a}{Number of data points in light curve. \textsuperscript{b}Total exposure time (s). \textsuperscript{c}Percent of rotation period covered. \textsuperscript{d}Read noise (e\textsuperscript{-}). \textsuperscript{e}Gain (e\textsuperscript{-}/ADU). }
\end{deluxetable}

\clearpage
\begin{deluxetable}{p{2.0 cm}cccccl}
\tabletypesize{\scriptsize}
\tablewidth{0pt}
\tablecaption{(3200) Phaethon Surface Color Variation \label{tab-color}}
\tablehead{
   \colhead{UT}
 & \colhead{True Anomaly ($^{\circ}$)}
 & \colhead{Sub-Earth Latitude ($^{\circ}$)}
 & \colhead{$B-V$}
 & \colhead{$V-R$}
 & \colhead{Reference}
}
\startdata
2010/09/10 & 201 & --62 & 0.67 $\pm$ 0.02 & 0.32 $\pm$ 0.02 & \cite{2013AJ....145..133J} \\
2007/09/04 & 193 & --62 & 0.61 $\pm$ 0.01 & 0.34 $\pm$ 0.03 & \cite{2008AJ....136..881K} \\
2004/11/19 & 197 & --50 & 0.59 $\pm$ 0.01 & 0.35 $\pm$ 0.01 & \cite{2004IfA} \\
1997/11/12 & 210 & --40 & 0.58 $\pm$ 0.01 & 0.34 $\pm$ 0.02 & This work \\
1997/11/22 & 214 & --22 & 0.57 $\pm$ 0.01 & 0.36 $\pm$ 0.01 & This work \\
1995/01/04 & 211 & --11 & 0.52 $\pm$ 0.01 & 0.33 $\pm$ 0.01 & This work \\
\enddata
\end{deluxetable}%

\clearpage
\begin{deluxetable}{p{2.0 cm}ccc}
\tabletypesize{\scriptsize}
\tablewidth{0pt}
\tablecaption{Observed Light Curves of (3200) Phaethon\textsuperscript{a} \label{tab-lcs}}
\tablehead{
   \colhead{JD}
 & \colhead{UT}
 & \colhead{{\it R}}
 & \colhead{$\sigma_{\it R}$}
}
\startdata
2449713.5767  &  1.8418  &  14.452  &  0.005 \\
2449713.5817  &  1.9598  &  14.477  &  0.005 \\
2449713.5870  &  2.0878  &  14.531  &  0.005 \\
2449713.5917  &  2.2020  &  14.554  &  0.005 \\
2449713.5965  &  2.3170  &  14.547  &  0.005 \\
2449713.6079  &  2.5906  &  14.496  &  0.005 \\
2449713.6101  &  2.6431  &  14.474  &  0.005 \\
2449713.6123  &  2.6959  &  14.474  &  0.005 \\
2449713.6145  &  2.7487  &  14.460  &  0.005 \\
2449713.6167  &  2.8018  &  14.438  &  0.005 \\
2449713.6189  &  2.8541  &  14.425  &  0.005 \\
2449713.6211  &  2.9069  &  14.414  &  0.005 \\
2449713.6277  &  3.0653  &  14.391  &  0.005 \\
2449713.6299  &  3.1183  &  14.391  &  0.005 \\
2449713.6321  &  3.1711  &  14.398  &  0.005 \\
2449713.6343  &  3.2239  &  14.397  &  0.005 \\
2449713.6365  &  3.2765  &  14.391  &  0.005 \\
2449713.6387  &  3.3295  &  14.396  &  0.005 \\
2449713.6409  &  3.3818  &  14.402  &  0.005 \\
2449713.6431  &  3.4346  &  14.400  &  0.005 \\
2449713.6453  &  3.4874  &  14.414  &  0.005 \\
2449713.6475  &  3.5405  &  14.440  &  0.005 \\
2449713.6499  &  3.5969  &  14.464  &  0.005 \\
2449713.6521  &  3.6497  &  14.494  &  0.005 \\
2449713.6543  &  3.7027  &  14.518  &  0.005 \\
2449713.6565  &  3.7555  &  14.537  &  0.005 \\
2449713.6587  &  3.8081  &  14.553  &  0.005 \\
2449713.6609  &  3.8609  &  14.564  &  0.005 \\
2449713.6631  &  3.9137  &  14.572  &  0.005 \\
2449713.6653  &  3.9667  &  14.584  &  0.005 \\
2449713.6675  &  4.0195  &  14.587  &  0.005 \\
2449713.6781  &  4.2737  &  14.543  &  0.005 \\
2449713.6810  &  4.3440  &  14.540  &  0.005 \\
2449713.6832  &  4.3966  &  14.542  &  0.005 \\
2449713.6854  &  4.4494  &  14.537  &  0.005 \\
2449713.6876  &  4.5024  &  14.530  &  0.005 \\
2449713.6898  &  4.5550  &  14.526  &  0.005 \\
2449713.6920  &  4.6078  &  14.538  &  0.006 \\
2449713.6942  &  4.6606  &  14.532  &  0.006 \\
2449713.6964  &  4.7134  &  14.520  &  0.006 \\
2449713.6986  &  4.7662  &  14.510  &  0.006 \\
2449713.7008  &  4.8190  &  14.490  &  0.006 \\
2449713.7030  &  4.8718  &  14.467  &  0.006 \\
2449713.7052  &  4.9246  &  14.463  &  0.006 \\
2449713.7074  &  4.9771  &  14.444  &  0.006 \\
2449713.7096  &  5.0299  &  14.431  &  0.006 \\
2449713.7118  &  5.0827  &  14.426  &  0.006 \\
2449713.7140  &  5.1358  &  14.431  &  0.006 \\
2449713.7162  &  5.1883  &  14.437  &  0.006 \\
2449713.7184  &  5.2411  &  14.440  &  0.006 \\
2449713.7206  &  5.2939  &  14.440  &  0.006 \\
2449713.7228  &  5.3467  &  14.445  &  0.006 \\
2449713.7395  &  5.7482  &  14.572  &  0.007 \\
2449713.7419  &  5.8054  &  14.570  &  0.007 \\
2449713.7441  &  5.8582  &  14.565  &  0.007 \\
2449713.7463  &  5.9107  &  14.568  &  0.007 \\
2449713.7485  &  5.9635  &  14.567  &  0.007 \\
2449713.7507  &  6.0163  &  14.558  &  0.007 \\
2449713.7529  &  6.0691  &  14.552  &  0.007 \\
2449713.7551  &  6.1219  &  14.536  &  0.007 \\
2449713.7573  &  6.1747  &  14.528  &  0.007 \\
2449713.7595  &  6.2275  &  14.514  &  0.007 \\
2449713.7617  &  6.2806  &  14.493  &  0.007 \\
2449713.7639  &  6.3331  &  14.477  &  0.007 \\
2449713.7661  &  6.3859  &  14.458  &  0.007 \\
2449713.7683  &  6.4387  &  14.448  &  0.007 \\
2449713.7705  &  6.4918  &  14.433  &  0.008 \\
2449713.7727  &  6.5441  &  14.424  &  0.008 \\
2449713.7749  &  6.5969  &  14.416  &  0.008 \\
2449713.7771  &  6.6497  &  14.411  &  0.008 \\
2449713.7793  &  6.7025  &  14.404  &  0.008 \\
2449713.7860  &  6.8647  &  14.408  &  0.009 \\
2449713.7882  &  6.9178  &  14.414  &  0.009 \\
2449713.7904  &  6.9703  &  14.418  &  0.009 \\
2449713.7926  &  7.0231  &  14.426  &  0.009 \\
2449713.7994  &  7.1866  &  14.490  &  0.009 \\
2449721.8667  &  8.8019  &  15.032  &  0.005 \\
2449721.8684  &  8.8414  &  15.022  &  0.005 \\
2449721.8702  &  8.8839  &  14.983  &  0.005 \\
2449721.8719  &  8.9244  &  14.955  &  0.005 \\
2449721.8735  &  8.9642  &  14.929  &  0.005 \\
2449721.8752  &  9.0047  &  14.907  &  0.005 \\
2449721.8819  &  9.1664  &  14.832  &  0.005 \\
2449721.8836  &  9.2067  &  14.821  &  0.005 \\
2449721.8853  &  9.2472  &  14.816  &  0.005 \\
2449721.8870  &  9.2872  &  14.796  &  0.006 \\
2449721.8886  &  9.3275  &  14.792  &  0.006 \\
2449722.7206  &  5.2953  &  15.030  &  0.002 \\
2449722.7220  &  5.3269  &  15.023  &  0.002 \\
2449722.7234  &  5.3611  &  15.016  &  0.002 \\
2449722.7274  &  5.4581  &  15.022  &  0.002 \\
2449722.7297  &  5.5128  &  15.026  &  0.002 \\
2449722.7307  &  5.5367  &  15.023  &  0.002 \\
2449722.7331  &  5.5933  &  15.024  &  0.002 \\
2449722.7348  &  5.6353  &  15.030  &  0.002 \\
2449722.7365  &  5.6758  &  15.040  &  0.002 \\
2449722.7383  &  5.7194  &  15.056  &  0.002 \\
2449722.7400  &  5.7594  &  15.069  &  0.002 \\
2449722.7417  &  5.8014  &  15.086  &  0.003 \\
2449722.7453  &  5.8869  &  15.138  &  0.002 \\
2449722.7487  &  5.9686  &  15.165  &  0.002 \\
2449722.7504  &  6.0086  &  15.162  &  0.002 \\
2449722.7521  &  6.0494  &  15.166  &  0.002 \\
2449722.7537  &  6.0900  &  15.169  &  0.002 \\
2449722.7554  &  6.1294  &  15.181  &  0.002 \\
2449722.7570  &  6.1686  &  15.193  &  0.002 \\
2449722.7587  &  6.2086  &  15.193  &  0.002 \\
2449722.7604  &  6.2492  &  15.192  &  0.002 \\
2449722.7622  &  6.2931  &  15.188  &  0.002 \\
2449722.7639  &  6.3331  &  15.175  &  0.002 \\
2449722.7655  &  6.3731  &  15.159  &  0.002 \\
2449722.7672  &  6.4131  &  15.132  &  0.002 \\
2449722.7690  &  6.4558  &  15.108  &  0.002 \\
2449722.7707  &  6.4961  &  15.089  &  0.002 \\
2449722.7723  &  6.5361  &  15.065  &  0.002 \\
2449722.7741  &  6.5778  &  15.034  &  0.002 \\
2449722.7781  &  6.6756  &  14.988  &  0.002 \\
2449722.7835  &  6.8047  &  14.927  &  0.002 \\
2449722.7852  &  6.8458  &  14.897  &  0.002 \\
2449722.7869  &  6.8867  &  14.894  &  0.002 \\
2449722.7886  &  6.9272  &  14.889  &  0.002 \\
2449722.7903  &  6.9672  &  14.887  &  0.002 \\
2449722.7920  &  7.0072  &  14.889  &  0.002 \\
2449722.7936  &  7.0469  &  14.907  &  0.002 \\
2449722.7953  &  7.0883  &  14.916  &  0.002 \\
2449722.7972  &  7.1317  &  14.939  &  0.002 \\
2449722.7989  &  7.1725  &  14.962  &  0.002 \\
2449722.8006  &  7.2133  &  14.979  &  0.002 \\
2449722.8024  &  7.2581  &  15.001  &  0.002 \\
2449722.8043  &  7.3036  &  15.014  &  0.002 \\
2449722.8060  &  7.3444  &  15.023  &  0.003 \\
2449722.8094  &  7.4264  &  15.062  &  0.003 \\
2449722.8129  &  7.5100  &  15.100  &  0.003 \\
2449722.8146  &  7.5500  &  15.122  &  0.003 \\
2449722.8163  &  7.5914  &  15.143  &  0.003 \\
2449722.8182  &  7.6364  &  15.161  &  0.003 \\
2449722.8200  &  7.6800  &  15.156  &  0.003 \\
2449722.8217  &  7.7214  &  15.166  &  0.003 \\
2449722.8234  &  7.7625  &  15.162  &  0.003 \\
2449722.8252  &  7.8047  &  15.166  &  0.003 \\
2449722.8270  &  7.8486  &  15.167  &  0.003 \\
2449722.8289  &  7.8947  &  15.162  &  0.003 \\
2449722.8308  &  7.9394  &  15.152  &  0.003 \\
2449722.8325  &  7.9800  &  15.147  &  0.003 \\
2449722.8343  &  8.0225  &  15.151  &  0.003 \\
2449722.8360  &  8.0639  &  15.154  &  0.003 \\
2449722.8378  &  8.1061  &  15.154  &  0.003 \\
2449722.8395  &  8.1483  &  15.166  &  0.003 \\
2449722.8413  &  8.1906  &  15.171  &  0.003 \\
2449722.8433  &  8.2403  &  15.179  &  0.003 \\
2449722.8453  &  8.2883  &  15.194  &  0.003 \\
2449722.8472  &  8.3317  &  15.212  &  0.003 \\
2449722.8489  &  8.3747  &  15.219  &  0.003 \\
2449722.8508  &  8.4181  &  15.227  &  0.003 \\
2449722.8526  &  8.4614  &  15.207  &  0.003 \\
2449722.8544  &  8.5047  &  15.182  &  0.003 \\
2449722.8561  &  8.5456  &  15.162  &  0.003 \\
2449722.8578  &  8.5869  &  15.125  &  0.003 \\
2449722.8595  &  8.6289  &  15.111  &  0.003 \\
2449722.8625  &  8.6997  &  15.067  &  0.003 \\
2449722.8661  &  8.7856  &  15.043  &  0.003 \\
2449722.8678  &  8.8264  &  15.021  &  0.003 \\
2449722.8697  &  8.8717  &  15.013  &  0.003 \\
2449722.8712  &  8.9094  &  15.004  &  0.003 \\
2449722.8732  &  8.9558  &  15.015  &  0.003 \\
2449722.8751  &  9.0017  &  15.009  &  0.003 \\
2450764.0178  & 12.4283  &  15.219  &  0.004 \\
2450764.0218  & 12.5228  &  15.230  &  0.004 \\
2450764.0254  & 12.6092  &  15.228  &  0.004 \\
2450764.0278  & 12.6667  &  15.228  &  0.004 \\
2450764.0302  & 12.7244  &  15.231  &  0.004 \\
2450764.0325  & 12.7806  &  15.233  &  0.004 \\
2450764.0353  & 12.8475  &  15.233  &  0.004 \\
2450764.0375  & 12.9011  &  15.235  &  0.004 \\
2450764.0398  & 12.9558  &  15.236  &  0.004 \\
2450764.0598  & 13.4364  &  15.302  &  0.003 \\
2450764.0628  & 13.5081  &  15.298  &  0.003 \\
2450764.0652  & 13.5656  &  15.290  &  0.003 \\
2450764.0675  & 13.6208  &  15.290  &  0.003 \\
2450764.0699  & 13.6783  &  15.289  &  0.003 \\
2450764.0725  & 13.7408  &  15.280  &  0.003 \\
2450764.0750  & 13.8003  &  15.282  &  0.003 \\
2450764.0780  & 13.8711  &  15.284  &  0.003 \\
2450764.0802  & 13.9247  &  15.278  &  0.003 \\
2450764.0831  & 13.9933  &  15.268  &  0.003 \\
2450764.0858  & 14.0594  &  15.261  &  0.003 \\
2450764.0885  & 14.1242  &  15.251  &  0.003 \\
2450764.0912  & 14.1883  &  15.236  &  0.003 \\
2450764.0941  & 14.2594  &  15.219  &  0.003 \\
2450764.0969  & 14.3244  &  15.209  &  0.003 \\
2450764.0995  & 14.3878  &  15.202  &  0.003 \\
2450764.1021  & 14.4497  &  15.196  &  0.003 \\
2450764.1061  & 14.5467  &  15.185  &  0.003 \\
2450764.1086  & 14.6075  &  15.176  &  0.003 \\
2450764.1117  & 14.6803  &  15.168  &  0.003 \\
2450764.1140  & 14.7358  &  15.166  &  0.003 \\
2450764.1162  & 14.7900  &  15.168  &  0.003 \\
2450764.1188  & 14.8512  &  15.172  &  0.003 \\
2450764.1219  & 14.9256  &  15.176  &  0.003 \\
2450764.1244  & 14.9864  &  15.169  &  0.003 \\
2450764.1276  & 15.0622  &  15.163  &  0.003 \\
2450764.1302  & 15.1258  &  15.162  &  0.003 \\
2450764.1329  & 15.1892  &  15.160  &  0.003 \\
2450764.1446  & 15.4700  &  15.147  &  0.003 \\
2450764.1467  & 15.5214  &  15.146  &  0.003 \\
2450765.0048  & 12.1161  &  15.123  &  0.004 \\
2450765.0070  & 12.1672  &  15.117  &  0.003 \\
2450765.0088  & 12.2117  &  15.110  &  0.003 \\
2450765.0107  & 12.2558  &  15.108  &  0.003 \\
2450765.0125  & 12.3006  &  15.107  &  0.003 \\
2450765.0144  & 12.3464  &  15.109  &  0.003 \\
2450765.0164  & 12.3928  &  15.113  &  0.003 \\
2450765.0183  & 12.4392  &  15.114  &  0.003 \\
2450765.0202  & 12.4847  &  15.122  &  0.003 \\
2450765.0276  & 12.6625  &  15.102  &  0.003 \\
2450765.0294  & 12.7067  &  15.098  &  0.003 \\
2450765.0313  & 12.7503  &  15.097  &  0.003 \\
2450765.0331  & 12.7936  &  15.096  &  0.003 \\
2450765.0360  & 12.8647  &  15.093  &  0.003 \\
2450765.0381  & 12.9142  &  15.092  &  0.003 \\
2450765.0399  & 12.9575  &  15.088  &  0.003 \\
2450765.0417  & 13.0019  &  15.088  &  0.003 \\
2450765.0436  & 13.0453  &  15.088  &  0.003 \\
2450765.0454  & 13.0906  &  15.090  &  0.003 \\
2450765.0473  & 13.1353  &  15.089  &  0.003 \\
2450765.0492  & 13.1800  &  15.091  &  0.003 \\
2450765.0510  & 13.2244  &  15.093  &  0.003 \\
2450765.0529  & 13.2689  &  15.094  &  0.003 \\
2450765.0548  & 13.3150  &  15.103  &  0.003 \\
2450765.0571  & 13.3711  &  15.113  &  0.003 \\
2450765.0595  & 13.4286  &  15.118  &  0.003 \\
2450765.0615  & 13.4750  &  15.128  &  0.003 \\
2450765.0634  & 13.5214  &  15.141  &  0.003 \\
2450765.0656  & 13.5742  &  15.145  &  0.003 \\
2450765.0689  & 13.6539  &  15.155  &  0.003 \\
2450765.0714  & 13.7136  &  15.164  &  0.003 \\
2450765.0743  & 13.7833  &  15.162  &  0.003 \\
2450765.0765  & 13.8356  &  15.159  &  0.003 \\
2450765.0787  & 13.8894  &  15.162  &  0.003 \\
2450765.0815  & 13.9564  &  15.164  &  0.003 \\
2450765.0859  & 14.0628  &  15.167  &  0.003 \\
2450765.0878  & 14.1078  &  15.165  &  0.003 \\
2450765.0927  & 14.2242  &  15.174  &  0.003 \\
2450765.0945  & 14.2692  &  15.179  &  0.003 \\
2450765.0964  & 14.3131  &  15.186  &  0.003 \\
2450765.1051  & 14.5231  &  15.208  &  0.003 \\
2450765.1070  & 14.5686  &  15.216  &  0.003 \\
2450765.1089  & 14.6133  &  15.222  &  0.003 \\
2450765.1107  & 14.6575  &  15.224  &  0.003 \\
2450765.1312  & 15.1483  &  15.204  &  0.003 \\
2450765.1330  & 15.1925  &  15.199  &  0.003 \\
2450765.1348  & 15.2356  &  15.191  &  0.003 \\
2450765.1367  & 15.2803  &  15.182  &  0.003 \\
2450765.1385  & 15.3239  &  15.175  &  0.003 \\
2450765.1403  & 15.3669  &  15.170  &  0.003 \\
2450765.1421  & 15.4106  &  15.155  &  0.003 \\
2450765.1440  & 15.4558  &  15.143  &  0.003 \\
2450773.9918  & 11.8039  &  14.743  &  0.005 \\
2450773.9972  & 11.9328  &  14.692  &  0.004 \\
2450773.9999  & 11.9972  &  14.652  &  0.004 \\
2450774.0025  & 12.0608  &  14.636  &  0.004 \\
2450774.0052  & 12.1236  &  14.639  &  0.004 \\
2450774.0078  & 12.1869  &  14.627  &  0.004 \\
2450774.0104  & 12.2500  &  14.614  &  0.004 \\
2450774.0131  & 12.3133  &  14.617  &  0.004 \\
2450774.0201  & 12.4825  &  14.595  &  0.004 \\
2450774.0213  & 12.5119  &  14.594  &  0.004 \\
2450774.0230  & 12.5511  &  14.602  &  0.004 \\
2450774.0241  & 12.5792  &  14.621  &  0.004 \\
2450774.0253  & 12.6061  &  14.610  &  0.004 \\
2450774.0264  & 12.6342  &  14.612  &  0.004 \\
2450774.0276  & 12.6622  &  14.614  &  0.004 \\
2450774.0288  & 12.6917  &  14.614  &  0.004 \\
2450774.0300  & 12.7189  &  14.610  &  0.004 \\
2450774.0311  & 12.7467  &  14.611  &  0.004 \\
2450774.0323  & 12.7753  &  14.609  &  0.004 \\
2450774.0335  & 12.8039  &  14.602  &  0.004 \\
2450774.0346  & 12.8314  &  14.604  &  0.004 \\
2450774.0358  & 12.8589  &  14.602  &  0.004 \\
2450774.0371  & 12.8900  &  14.596  &  0.003 \\
2450774.0383  & 12.9194  &  14.602  &  0.003 \\
2450774.0395  & 12.9478  &  14.593  &  0.003 \\
2450774.0408  & 12.9786  &  14.594  &  0.003 \\
2450774.0419  & 13.0064  &  14.585  &  0.003 \\
2450774.0431  & 13.0344  &  14.583  &  0.003 \\
2450774.0443  & 13.0625  &  14.574  &  0.003 \\
2450774.0455  & 13.0908  &  14.573  &  0.003 \\
2450774.0467  & 13.1208  &  14.572  &  0.003 \\
2450774.0478  & 13.1481  &  14.566  &  0.003 \\
2450774.0490  & 13.1753  &  14.567  &  0.003 \\
2450774.0501  & 13.2028  &  14.566  &  0.003 \\
2450774.0574  & 13.3781  &  14.580  &  0.003 \\
2450774.0587  & 13.4081  &  14.584  &  0.003 \\
2450774.0598  & 13.4353  &  14.590  &  0.003 \\
2450774.0610  & 13.4650  &  14.590  &  0.003 \\
2450774.0630  & 13.5128  &  14.608  &  0.003 \\
2450774.0642  & 13.5408  &  14.616  &  0.003 \\
2450774.0656  & 13.5744  &  14.630  &  0.003 \\
2450774.0667  & 13.6019  &  14.640  &  0.003 \\
2450774.0679  & 13.6306  &  14.646  &  0.003 \\
2450774.0691  & 13.6594  &  14.660  &  0.003 \\
2450774.0704  & 13.6889  &  14.679  &  0.003 \\
2450774.0715  & 13.7167  &  14.682  &  0.003 \\
2450774.0727  & 13.7436  &  14.691  &  0.003 \\
2450774.0738  & 13.7714  &  14.699  &  0.003 \\
2450775.0445  & 13.0683  &  14.642  &  0.003 \\
2450775.0467  & 13.1200  &  14.612  &  0.004 \\
2450775.0484  & 13.1617  &  14.591  &  0.004 \\
2450775.0500  & 13.2008  &  14.584  &  0.004 \\
2450775.0520  & 13.2469  &  14.569  &  0.004 \\
2450775.0542  & 13.3019  &  14.558  &  0.004 \\
2450775.0560  & 13.3431  &  14.574  &  0.004 \\
2450775.0582  & 13.3978  &  14.568  &  0.004 \\
2450775.0591  & 13.4175  &  14.558  &  0.004 \\
2450775.0598  & 13.4361  &  14.560  &  0.004 \\
2450775.0606  & 13.4550  &  14.562  &  0.004 \\
2450775.0614  & 13.4742  &  14.564  &  0.004 \\
2450775.0622  & 13.4928  &  14.567  &  0.004 \\
2450775.0630  & 13.5125  &  14.565  &  0.004 \\
2450775.0665  & 13.5956  &  14.559  &  0.004 \\
2450775.0673  & 13.6144  &  14.557  &  0.004 \\
2450775.0681  & 13.6342  &  14.561  &  0.004 \\
2450775.0689  & 13.6536  &  14.560  &  0.004 \\
2450775.0697  & 13.6728  &  14.566  &  0.004 \\
2450775.0705  & 13.6931  &  14.562  &  0.004 \\
2450775.0737  & 13.7692  &  14.571  &  0.004 \\
2450775.0769  & 13.8461  &  14.579  &  0.004 \\
2450775.0801  & 13.9233  &  14.583  &  0.004 \\
2450775.0850  & 14.0392  &  14.575  &  0.004 \\
2450775.0881  & 14.1133  &  14.576  &  0.004 \\
2450775.0914  & 14.1939  &  14.559  &  0.004 \\
2450775.0946  & 14.2708  &  14.553  &  0.004 \\
2450775.0954  & 14.2903  &  14.549  &  0.004 \\
2450775.0989  & 14.3731  &  14.539  &  0.004 \\
2450775.1015  & 14.4362  &  14.546  &  0.005 \\
2450775.1064  & 14.5539  &  14.536  &  0.005 \\
2450775.1088  & 14.6111  &  14.551  &  0.005 \\
2450775.1114  & 14.6747  &  14.567  &  0.005 \\
2450775.1136  & 14.7275  &  14.569  &  0.005 \\
2450775.1159  & 14.7808  &  14.605  &  0.005 \\
2450775.1181  & 14.8353  &  14.614  &  0.005 \\
2450775.1207  & 14.8978  &  14.660  &  0.005 \\
2450775.1232  & 14.9567  &  14.689  &  0.006 \\
2450775.1271  & 15.0514  &  14.699  &  0.006 \\
2450775.1277  & 15.0642  &  14.688  &  0.006 \\
2450775.1283  & 15.0781  &  14.689  &  0.006 \\
2450778.0237  & 12.5683  &  14.835  &  0.015 \\
2450778.0250  & 12.6003  &  14.836  &  0.015 \\
2450778.0264  & 12.6344  &  14.832  &  0.015 \\
2450778.0279  & 12.6706  &  14.834  &  0.014 \\
2450778.0306  & 12.7350  &  14.816  &  0.014 \\
2450778.0315  & 12.7558  &  14.805  &  0.014 \\
2450778.0331  & 12.7944  &  14.788  &  0.013 \\
2450778.0339  & 12.8144  &  14.773  &  0.013 \\
2450778.0348  & 12.8347  &  14.769  &  0.013 \\
2450778.0359  & 12.8608  &  14.756  &  0.013 \\
2450778.0367  & 12.8800  &  14.747  &  0.013 \\
2450778.0379  & 12.9086  &  14.737  &  0.013 \\
2450778.0390  & 12.9369  &  14.731  &  0.013 \\
2450778.0412  & 12.9900  &  14.703  &  0.012 \\
2450778.0423  & 13.0144  &  14.680  &  0.012 \\
2450778.0439  & 13.0542  &  14.668  &  0.012 \\
2450778.0448  & 13.0764  &  14.652  &  0.012 \\
2450778.0460  & 13.1047  &  14.641  &  0.012 \\
2450778.0470  & 13.1289  &  14.636  &  0.012 \\
2450778.0478  & 13.1478  &  14.631  &  0.012 \\
2450778.0487  & 13.1683  &  14.620  &  0.012 \\
2450778.0505  & 13.2114  &  14.613  &  0.011 \\
2450778.0513  & 13.2311  &  14.604  &  0.011 \\
2450778.0521  & 13.2503  &  14.595  &  0.011 \\
2453328.8286  &  7.8878  &  15.773  &  0.005 \\
2453328.8308  &  7.9406  &  15.775  &  0.005 \\
2453328.8391  &  8.1367  &  15.749  &  0.005 \\
2453328.8457  &  8.2989  &  15.722  &  0.005 \\
2453328.8494  &  8.3867  &  15.700  &  0.005 \\
2453328.8530  &  8.4733  &  15.686  &  0.005 \\
2453328.8567  &  8.5597  &  15.671  &  0.005 \\
2453328.8601  &  8.6436  &  15.666  &  0.005 \\
2453328.8638  &  8.7308  &  15.670  &  0.005 \\
2453328.8672  &  8.8117  &  15.679  &  0.005 \\
2453328.8706  &  8.8922  &  15.687  &  0.005 \\
2453328.8740  &  8.9758  &  15.715  &  0.005 \\
2453328.8806  &  9.1378  &  15.761  &  0.005 \\
2453328.8840  &  9.2183  &  15.777  &  0.005 \\
2453328.8884  &  9.3219  &  15.785  &  0.005 \\
2453328.8918  &  9.4056  &  15.756  &  0.005 \\
2453328.8953  &  9.4864  &  15.743  &  0.005 \\
2453328.8987  &  9.5672  &  15.727  &  0.005 \\
2453328.9021  &  9.6481  &  15.717  &  0.005 \\
2453328.9060  &  9.7411  &  15.715  &  0.005 \\
2453328.9141  &  9.9383  &  15.696  &  0.005 \\
2453328.9175  & 10.0194  &  15.702  &  0.005 \\
2453328.9209  & 10.1003  &  15.715  &  0.005 \\
2453328.9248  & 10.1931  &  15.708  &  0.005 \\
2453328.9280  & 10.2736  &  15.704  &  0.005 \\
2453328.9446  & 10.6692  &  15.696  &  0.005 \\
2453328.9480  & 10.7500  &  15.710  &  0.005 \\
2453328.9570  & 10.9703  &  15.715  &  0.005 \\
2453328.9604  & 11.0511  &  15.723  &  0.005 \\
2453328.9639  & 11.1325  &  15.750  &  0.005 \\
2453328.9673  & 11.2161  &  15.749  &  0.005 \\
2453328.9768  & 11.4422  &  15.742  &  0.005 \\
2453328.9834  & 11.6028  &  15.759  &  0.005 \\
2453328.9854  & 11.6489  &  15.752  &  0.005 \\
2453328.9871  & 11.6894  &  15.750  &  0.005 \\
2453328.9888  & 11.7297  &  15.743  &  0.005 \\
2453328.9905  & 11.7708  &  15.734  &  0.005 \\
2453328.9922  & 11.8111  &  15.741  &  0.005 \\
2453330.7798  &  6.7144  &  15.541  &  0.006 \\
2453330.7833  &  6.7986  &  15.555  &  0.006 \\
2453330.7869  &  6.8847  &  15.533  &  0.006 \\
2453330.7902  &  6.9653  &  15.525  &  0.006 \\
2453330.7970  &  7.1269  &  15.504  &  0.007 \\
2453330.8007  &  7.2175  &  15.473  &  0.006 \\
2453330.8041  &  7.2986  &  15.452  &  0.007 \\
2453330.8075  &  7.3792  &  15.443  &  0.005 \\
2453330.8108  &  7.4603  &  15.439  &  0.007 \\
2453330.8142  &  7.5408  &  15.437  &  0.006 \\
2453330.8844  &  9.2251  &  15.461  &  0.004 \\
2453330.9140  &  9.9364  &  15.503  &  0.004 \\
2453330.9174  & 10.0183  &  15.510  &  0.004 \\
2453330.9244  & 10.1864  &  15.520  &  0.004 \\
2453330.9278  & 10.2669  &  15.515  &  0.004 \\
2453330.9311  & 10.3475  &  15.525  &  0.004 \\
2453330.9346  & 10.4283  &  15.528  &  0.005 \\
2453330.9380  & 10.5092  &  15.528  &  0.005 \\
2453330.9446  & 10.6708  &  15.503  &  0.004 \\
2453330.9487  & 10.7678  &  15.481  &  0.005 \\
2453330.9548  & 10.9144  &  15.442  &  0.005 \\
2453330.9583  & 10.9961  &  15.434  &  0.004 \\
2453330.9636  & 11.1264  &  15.422  &  0.004 \\
2453330.9670  & 11.2072  &  15.424  &  0.004 \\
2453330.9734  & 11.3647  &  15.453  &  0.004 \\
2453330.9785  & 11.4861  &  15.483  &  0.004 \\
2453330.9834  & 11.6017  &  15.513  &  0.004 \\
2453330.9868  & 11.6822  &  15.544  &  0.004 \\
2453330.9902  & 11.7675  &  15.539  &  0.004 \\
2453330.9919  & 11.8078  &  15.537  &  0.004 \\
2453330.9966  & 11.9156  &  15.520  &  0.004 \\
2453331.0066  & 12.1617  &  15.467  &  0.004 \\
2453331.0100  & 12.2422  &  15.461  &  0.004 \\
2453331.0122  & 12.2964  &  15.455  &  0.004 \\
2453331.0173  & 12.4172  &  15.461  &  0.004 \\
2453331.0190  & 12.4575  &  15.457  &  0.004 \\
2453331.0227  & 12.5417  &  15.469  &  0.004 \\
2453331.0254  & 12.6061  &  15.468  &  0.004 \\
2453331.0327  & 12.7839  &  15.463  &  0.004 \\
2453331.0388  & 12.9289  &  15.459  &  0.004 \\
2453331.0422  & 13.0144  &  15.453  &  0.004 \\
2453331.0466  & 13.1186  &  15.461  &  0.004 \\
2453331.0503  & 13.2058  &  15.479  &  0.004 \\
2453331.0520  & 13.2461  &  15.466  &  0.004 \\
2453331.0540  & 13.2942  &  15.470  &  0.004 \\
2453331.0557  & 13.3367  &  15.469  &  0.004 \\
2453331.0591  & 13.4172  &  15.477  &  0.004 \\
2453331.0625  & 13.4978  &  15.481  &  0.004 \\
2453331.0642  & 13.5383  &  15.500  &  0.004 \\
2453331.0671  & 13.6114  &  15.510  &  0.004 \\
2453331.0688  & 13.6517  &  15.506  &  0.004 \\
2453331.7871  &  6.8908  &  15.419  &  0.005 \\
2453331.7888  &  6.9311  &  15.415  &  0.005 \\
2453331.7910  &  6.9842  &  15.411  &  0.005 \\
2453331.7927  &  7.0244  &  15.411  &  0.005 \\
2453331.7944  &  7.0647  &  15.408  &  0.005 \\
2453331.8047  &  7.3100  &  15.423  &  0.005 \\
2453331.8064  &  7.3503  &  15.421  &  0.005 \\
2453331.8079  &  7.3906  &  15.423  &  0.005 \\
2453331.8147  &  7.5517  &  15.451  &  0.005 \\
2453331.8164  &  7.5919  &  15.450  &  0.005 \\
2453331.8181  &  7.6322  &  15.451  &  0.005 \\
2453331.8247  &  7.7936  &  15.449  &  0.005 \\
2453331.8264  &  7.8339  &  15.452  &  0.005 \\
2453331.8281  &  7.8742  &  15.450  &  0.005 \\
2453331.8352  &  8.0414  &  15.461  &  0.005 \\
2453331.8367  &  8.0817  &  15.452  &  0.005 \\
2453331.8384  &  8.1219  &  15.455  &  0.004 \\
2453331.8452  &  8.2831  &  15.440  &  0.004 \\
2453331.8467  &  8.3233  &  15.430  &  0.004 \\
2453331.8484  &  8.3639  &  15.410  &  0.004 \\
2453331.8557  &  8.5383  &  15.371  &  0.004 \\
2453331.8574  &  8.5783  &  15.360  &  0.004 \\
2453331.8591  &  8.6186  &  15.354  &  0.004 \\
2453331.8660  &  8.7825  &  15.350  &  0.004 \\
2453331.8677  &  8.8231  &  15.345  &  0.004 \\
2453331.8711  &  8.9042  &  15.364  &  0.004 \\
2453331.8728  &  8.9444  &  15.367  &  0.004 \\
2453331.8745  &  8.9847  &  15.376  &  0.004 \\
2453331.8958  &  9.4964  &  15.451  &  0.004 \\
2453331.8975  &  9.5367  &  15.446  &  0.004 \\
2453331.8989  &  9.5769  &  15.437  &  0.004 \\
2453331.9058  &  9.7381  &  15.415  &  0.004 \\
2453331.9075  &  9.7783  &  15.409  &  0.004 \\
2453331.9092  &  9.8186  &  15.400  &  0.004 \\
2453331.9163  &  9.9908  &  15.386  &  0.004 \\
2456616.6899  &  4.5589  &  16.648  &  0.005 \\
2456616.6958  &  4.6961  &  16.657  &  0.006 \\
2456616.6997  &  4.7892  &  16.709  &  0.007 \\
2456616.7012  &  4.8339  &  16.734  &  0.007 \\
2456616.7031  &  4.8786  &  16.751  &  0.007 \\
2456616.7070  &  4.9700  &  16.773  &  0.008 \\
2456616.7090  &  5.0214  &  16.779  &  0.008 \\
2456616.7148  &  5.1575  &  16.799  &  0.008 \\
2456616.7168  &  5.2031  &  16.791  &  0.009 \\
2456616.7207  &  5.2956  &  16.780  &  0.008 \\
2456616.7271  &  5.4503  &  16.687  &  0.007 \\
2456616.7290  &  5.4958  &  16.641  &  0.006 \\
2456616.7329  &  5.5883  &  16.589  &  0.005 \\
2456616.7368  &  5.6811  &  16.519  &  0.005 \\
2456616.7427  &  5.8200  &  16.492  &  0.005 \\
2456616.7480  &  5.9564  &  16.467  &  0.007 \\
2456616.7500  &  6.0019  &  16.470  &  0.006 \\
2456616.7520  &  6.0494  &  16.480  &  0.009 \\
2456616.7559  &  6.1436  &  16.511  &  0.010 \\
2456616.7637  &  6.3247  &  16.524  &  0.006 \\
2456616.7720  &  6.5292  &  16.605  &  0.009 \\
2456616.7759  &  6.6239  &  16.637  &  0.009 \\
2456616.7798  &  6.7194  &  16.667  &  0.008 \\
2456616.7866  &  6.8803  &  16.732  &  0.011 \\
2456619.7528  &  6.0683  &  15.830  &  0.004 \\
2456619.7552  &  6.1258  &  15.835  &  0.004 \\
2456619.7596  &  6.2306  &  15.837  &  0.004 \\
2456619.7639  &  6.3333  &  15.836  &  0.004 \\
2456619.7662  &  6.3878  &  15.845  &  0.004 \\
2456619.7707  &  6.4975  &  15.864  &  0.004 \\
2456619.7729  &  6.5492  &  15.884  &  0.004 \\
2456619.7750  &  6.6003  &  15.905  &  0.005 \\
2456619.7773  &  6.6550  &  15.906  &  0.004 \\
2456619.7818  &  6.7625  &  15.970  &  0.004 \\
2456619.7840  &  6.8153  &  15.977  &  0.004 \\
2456619.7863  &  6.8711  &  16.012  &  0.005 \\
2456619.7948  &  7.0747  &  16.026  &  0.005 \\
2456619.7991  &  7.1781  &  16.004  &  0.005 \\
2456619.8012  &  7.2292  &  15.977  &  0.005 \\
2456619.8035  &  7.2833  &  15.972  &  0.005 \\
2456629.6040  &  2.4969  &  16.997  &  0.012 \\
2456629.6060  &  2.5414  &  17.019  &  0.015 \\
2456629.6074  &  2.5811  &  17.007  &  0.013 \\
2456629.6123  &  2.6986  &  17.048  &  0.012 \\
2456629.6143  &  2.7372  &  17.079  &  0.016 \\
2456629.6157  &  2.7761  &  17.093  &  0.012 \\
2456629.6172  &  2.8150  &  17.115  &  0.012 \\
2456629.6191  &  2.8542  &  17.115  &  0.011 \\
2456629.6211  &  2.9022  &  17.107  &  0.013 \\
2456629.6250  &  2.9964  &  17.118  &  0.014 \\
2456629.6270  &  3.0450  &  17.115  &  0.016 \\
2456629.6289  &  3.0917  &  17.082  &  0.017 \\
2456629.6309  &  3.1408  &  17.086  &  0.020 \\
2456629.6367  &  3.2761  &  17.018  &  0.017 \\
2456629.6426  &  3.4228  &  16.972  &  0.022 \\
2456629.6450  &  3.4828  &  16.923  &  0.018 \\
2456629.6699  &  4.0814  &  16.853  &  0.023 \\
2456629.6782  &  4.2714  &  16.918  &  0.019 \\
2456629.6821  &  4.3656  &  17.041  &  0.016 \\
2456629.6846  &  4.4328  &  17.047  &  0.013 \\
2456637.7035  &  4.8842  &  17.219  &  0.007 \\
2456637.7057  &  4.9372  &  17.227  &  0.007 \\
2456637.7083  &  4.9986  &  17.243  &  0.006 \\
2456637.7111  &  5.0672  &  17.281  &  0.007 \\
2456637.7137  &  5.1286  &  17.302  &  0.008 \\
2456637.7159  &  5.1814  &  17.309  &  0.008 \\
2456637.7181  &  5.2342  &  17.323  &  0.008 \\
2456637.7203  &  5.2872  &  17.331  &  0.009 \\
2456637.7225  &  5.3400  &  17.374  &  0.008 \\
2456637.7249  &  5.3975  &  17.381  &  0.009 \\
2456637.7283  &  5.4783  &  17.402  &  0.010 \\
2456637.7305  &  5.5319  &  17.398  &  0.009 \\
2456637.7327  &  5.5858  &  17.398  &  0.009 \\
2456637.7349  &  5.6386  &  17.390  &  0.009 \\
2456637.7384  &  5.7225  &  17.331  &  0.008 \\
2456637.7406  &  5.7753  &  17.307  &  0.010 \\
2456637.7429  &  5.8286  &  17.303  &  0.008 \\
2456637.7451  &  5.8819  &  17.289  &  0.008 \\
2456637.7473  &  5.9353  &  17.253  &  0.008 \\
2456637.7495  &  5.9883  &  17.234  &  0.008 \\
2456637.7517  &  6.0411  &  17.206  &  0.009 \\
2456637.7539  &  6.0939  &  17.202  &  0.008 \\
2456637.7561  &  6.1469  &  17.199  &  0.008 \\
2456637.7583  &  6.1997  &  17.175  &  0.008 \\
2456637.7605  &  6.2525  &  17.157  &  0.011 \\
2456637.7627  &  6.3056  &  17.155  &  0.009 \\
2456637.7649  &  6.3583  &  17.157  &  0.010 \\
2456637.7702  &  6.4850  &  17.151  &  0.006 \\
2456637.7730  &  6.5531  &  17.152  &  0.008 \\
2456637.7758  &  6.6203  &  17.156  &  0.007 \\
2456637.7786  &  6.6869  &  17.171  &  0.007 \\
2456637.7814  &  6.7536  &  17.170  &  0.008 \\
2456637.7842  &  6.8203  &  17.205  &  0.008 \\
2456637.7870  &  6.8872  &  17.224  &  0.008 \\
2456637.7930  &  7.0322  &  17.275  &  0.010 \\
2456637.7958  &  7.0994  &  17.303  &  0.010 \\
\enddata
\tablenotetext{a}{This table will be provided in Machine Readable Format in the online journal version of the paper.}

\end{deluxetable}

\end{document}